\documentclass[twocolumn]{aastex701}

\usepackage{amsmath}
\usepackage{bm}
\usepackage[dvipsnames]{xcolor}

\newcommand{\Cl}[1]{C_\ell^{#1}}
\newcommand{\dalpha}{\Delta\alpha}
\newcommand{\MK}{\mathrm{MK}}
\newcommand{\RelCal}{\mathrm{rel}}
\newcommand{\PTE}{\mathrm{PTE}}
\newcommand{\trans}{\mathsf{T}}
\newcommand{\npipe}{\texttt{NPIPE}}
\newcommand{\NaMaster}{\texttt{NaMaster}}
\providecommand{\degree}{}\renewcommand{\degree}{^\circ}

\shorttitle{Differential polarization calibration}
\shortauthors{Lonappan, Keating, \& Arnold}

\begin{document}

\title{Differential Polarization Calibration: A Consistency Test for Cosmic Birefringence}

\author[orcid=0000-0003-1200-9179,gname=Anto I.,sname=Lonappan]{Anto I. Lonappan}
\affiliation{Department of Physics, University of California, San Diego, CA 92093, USA}
\email[show]{alonappan@ucsd.edu}

\author[orcid=0000-0003-3118-5514,gname=Brian,sname=Keating]{Brian Keating}
\affiliation{Department of Physics, University of California, San Diego, CA 92093, USA}
\email{bkeating@ucsd.edu}

\author[orcid=0000-0002-3407-5305,gname=Kam,sname=Arnold]{Kam Arnold}
\affiliation{Department of Physics, University of California, San Diego, CA 92093, USA}
\affiliation{Department of Astronomy \& Astrophysics, University of California, San Diego, CA 92093, USA}
\email{karnold@ucsd.edu}

\correspondingauthor{Anto I. Lonappan}

\begin{abstract}
The cosmic microwave background birefringence angle $\beta$ is exactly degenerate with a common instrumental polarization-angle offset. The Minami--Komatsu(MK) likelihood separates these quantities using Galactic foreground polarization, so its inferred detector angles merit a diagnostic that does not reuse the foreground-$EB$ model. We develop a birefringence-blind differential estimator from antisymmetric and symmetric cross-spectra and apply it for the first time as an explicit consistency test of a foreground-assisted birefringence analysis. For a map pair, the common rotation cancels identically and the estimator measures $\Delta\alpha_{ij}=\alpha_i-\alpha_j$; combining all pairs in a gauge-fixed network reconstructs the observable detector-dependent calibration pattern. Applied to eight \textit{Planck} \npipe{} detector-set maps, the differential reconstruction agrees with the MK pattern, giving $\chi^2=8.18$ for seven degrees of freedom ($\PTE=0.32$), with all shared-data residuals below $1.8\sigma$. Because both reconstructions use the same maps, this is an independent estimator and modeling route rather than an independent-data confirmation. The agreement therefore strengthens confidence in the calibration foundation of the MK-inferred $\beta$ without providing an independent absolute measurement of it. A conditional common-mode reconstruction gives $\beta=0.37\pm0.12\degree$. The differential test uses existing data and can serve as a validation layer for future $EB$-based birefringence analyses.
\end{abstract}


\section{Introduction}\label{sec:intro}

Cosmic birefringence, a uniform rotation $\beta$ of the cosmic microwave background (CMB) polarization, would signal parity-violating physics beyond the Standard Model, including axion-like couplings to electromagnetism \citep{Carroll1990,Carroll1998,Lue1999,Feng2006,Komatsu2022review}. Recent analyses of \textit{Planck}, ACT and WMAP report $\beta\simeq0.3$--$0.4\degree$ at $2.4$--$4.8\sigma$ significance \citep{MinamiKomatsu2020b,DiegoPalazuelos2022,EskiltKomatsu2022,Eskilt:2026imm}. At this precision, instrumental polarization-angle calibration is the central systematic.

For map $i$, the CMB spectra depend on the total rotation $\theta_i=\alpha_i+\beta$, where $\alpha_i$ is the instrumental angle. CMB data alone therefore cannot separate the common sky rotation from instrumental offsets \citep{Wu2009,Komatsu2011,Keating2013}. The MK likelihood \citep{MinamiKomatsu2019,MinamiKomatsu2020b} breaks this degeneracy by using Galactic foreground polarization, which is rotated by $\alpha_i$ but not by $\beta$. This enables an absolute inference, but makes the result sensitive to the foreground $EB$ model, sky selection, and multipole range \citep{DiegoPalazuelos2022,Eskilt2022freq,Vacher2023,HerviasCaimapo2024}.

Here we develop a complementary differential calibration test and apply it for the first time as an explicit real-data consistency test of a foreground-assisted birefringence analysis. The idea is elementary: for any pair of maps the common rotation cancels, leaving $\theta_i-\theta_j=\alpha_i-\alpha_j$. We estimate this relative angle from antisymmetric and symmetric cross-spectrum combinations and combine all pairs in a gauge-fixed calibration network, in which the maps are nodes and the measured pairwise angle differences are edges. Solving this network reconstructs the detector-dependent pattern that is observable without fixing the common calibration mode. The test uses the same maps as the MK likelihood but a different estimator and a different foreground dependence, so it is an independent consistency check rather than an independent-data measurement of $\beta$.

The relative-angle algebra is related to the cross-experiment calibration method of \citet{LonappanSATxLAT2026}, which transfers calibration from a well-calibrated reference experiment. We address a different question: whether the detector-set angle pattern inferred by a foreground-assisted likelihood for a single experiment satisfies the corresponding differential network constraints in real \textit{Planck} data. Our contribution is therefore the gauge-explicit network comparison, its shared-data covariance, and its use as a validation layer for an absolute birefringence inference, not priority over the underlying identity.

Angle systematics also affect delensing and searches for inflationary $B$ modes through $E\to B$ mixing \citep{Abitbol2016,SimonsObservatory2019,LiteBIRD2023,IdicherianLonappan:2025trj}, and a differential check requires no additional observations, so it can accompany any absolute $EB$ analysis. We apply it to eight \textit{Planck} \npipe{} detector-set maps and compare with an independent implementation of the MK likelihood. Section~\ref{sec:method} gives the estimator and network formalism, Section~\ref{sec:data} describes the data, Section~\ref{sec:results} presents the consistency result, and Section~\ref{sec:anchor} briefly discusses the conditional absolute mode. Supporting derivations and sensitivity estimates are given in the Appendices.

\section{A Birefringence-Blind Differential Estimator}\label{sec:method}

\subsection{Relative-angle identity}\label{sec:relative}

Let $\bar E_{i,\ell m}$ and $\bar B_{i,\ell m}$ denote the pre-rotation (unrotated) harmonic-space fields of map $i$, including its beam and bandpass. The corresponding unbarred fields $E_{i,\ell m}$ and $B_{i,\ell m}$ are the observed fields after the total rotation. A rotation by $\theta_i=\alpha_i+\beta$ gives
\begin{equation}
\begin{pmatrix}E_i\\B_i\end{pmatrix}
=\begin{pmatrix}\cos2\theta_i&-\sin2\theta_i\\
\sin2\theta_i&\cos2\theta_i\end{pmatrix}
\begin{pmatrix}\bar E_i\\ \bar B_i\end{pmatrix}.
\end{equation}
Thus, a bar always denotes a pre-rotation quantity, while no bar denotes the observed, rotated quantity. For a map pair, define the observed combinations
\begin{equation}
D^{EB}_{ij,\ell}\equiv \Cl{B_iE_j}-\Cl{E_iB_j},
\qquad
S_{ij,\ell}\equiv \Cl{E_iE_j}+\Cl{B_iB_j}.
\end{equation}
For the corresponding pre-rotation combinations, write $\bar S_{ij,\ell}=\Cl{\bar E_i\bar E_j}+\Cl{\bar B_i\bar B_j}$ and $\bar D^{EB}_{ij,\ell}=\Cl{\bar B_i\bar E_j}-\Cl{\bar E_i\bar B_j}$. The rotation then gives exactly
\begin{align}
D^{EB}_{ij,\ell}&=\bar S_{ij,\ell}\sin[2(\theta_i-\theta_j)]+\bar D^{EB}_{ij,\ell}\cos[2(\theta_i-\theta_j)],
\label{eq:D_identity}\\
S_{ij,\ell}&=\bar S_{ij,\ell}\cos[2(\theta_i-\theta_j)]-\bar D^{EB}_{ij,\ell}\sin[2(\theta_i-\theta_j)].
\label{eq:S_identity}
\end{align}
Since $\theta_i-\theta_j=\alpha_i-\alpha_j\equiv\dalpha_{ij}$, $\beta$ cancels identically. If $\bar D^{EB}_{ij,\ell}=0$, then
\begin{equation}\label{eq:tan_relation}
D^{EB}_{ij,\ell}=\tan\!\big[2\dalpha_{ij}\big]S_{ij,\ell}.
\end{equation}

The condition $\bar D^{EB}_{ij,\ell}=0$ is exact for the CMB, even when a cosmological $\Cl{EB}$ is present, because the frequency-independent $E$ and $B$ fields acquire the symmetric beam product $b_i b_j$. For polarized foregrounds, $\bar D^{EB}_{ij}\propto[a_B(\nu_i)a_E(\nu_j)-a_E(\nu_i)a_B(\nu_j)]$. It therefore vanishes for same-frequency pairs for arbitrary SEDs and for cross-frequency pairs when the foreground $E$ and $B$ modes share a frequency scaling. We test the latter condition in Section~\ref{sec:data}; the remaining cross-frequency sensitivity is quantified in Appendix~\ref{app:relative}. The estimator does not use the foreground-$EB$ model that anchors the absolute MK likelihood.

\subsection{Pair likelihood and network reconstruction}\label{sec:pairlike}

For binned bandpowers $X_b=S_{ij,b}$ and $Y_b=D^{EB}_{ij,b}$, we estimate the pair angle by minimizing
\begin{equation}\label{eq:pair_chi2}
-2\ln\mathcal{L}_{ij}(\delta)=
\sum_b\frac{\big[Y_{b}\cos 2\delta-X_{b}\sin 2\delta\big]^2}
{\cos^2\!2\delta\,\mathrm{Var}(Y_{b})+\sin^2\!2\delta\,\mathrm{Var}(X_{b})},
\end{equation}
using Gaussian Knox-type variances \citep{Knox1995} from the measured spectra and mask. The $X$--$Y$ covariance is negligible at the measured parity-odd power; including it shifts the angles by much less than $0.1\sigma$. At small angles, $Y_b\simeq2\delta X_b$, so high-signal symmetric bandpowers carry the relative-angle information.

The $N_p=N(N-1)/2$ pair estimates are collected into the data vector $\bm d$, whose entries are the fitted pairwise differences $\hat{\dalpha}_{ij}$ from Equation~(\ref{eq:pair_chi2}), and they constrain the vector $\bm x$ of $N$ instrumental map angles $\alpha_i$ up to a common shift. The two are related by $\bm d=\bm A\bm x+\bm n$, where $\bm n$ is the noise on the pair estimates and $\bm A$ is the $N_p\times N$ graph incidence matrix: each row corresponds to one pair $(i,j)$ and has $+1$ in column $i$, $-1$ in column $j$, and zero elsewhere, so that $(\bm A\bm x)_{ij}=\alpha_i-\alpha_j$. Every row of $\bm A$ therefore sums to zero, and $\bm A\bm1=0$ exposes the common mode as the exact network gauge freedom. Fixing $\alpha_{143A}=0$ gives
\begin{equation}\label{eq:gls}
\hat{\bm x}_r=\big(\bm A_r^{\trans}\bm C_d^{-1}\bm A_r\big)^{-1}\bm A_r^{\trans}\bm C_d^{-1}\bm d,
\end{equation}
where the subscript $r$ denotes the gauge-reduced quantities: $\bm A_r$ is the incidence matrix with the column of the reference map removed, and $\hat{\bm x}_r$ is the corresponding vector of the remaining $N-1$ map angles, measured relative to that reference. Since $\bm A\bm x$ is deterministic, $\bm C_d=\langle\bm n\bm n^{\trans}\rangle$: the covariance of $\bm d$ and of $\bm n$ are the same object. We evaluate it by propagating the Gaussian bandpower covariance of Equation~(\ref{eq:pair_chi2}) through the pair fits. Its diagonal entries are the squared per-pair uncertainties $\sigma^2(\hat{\dalpha}_{ij})$, and its off-diagonal entries are non-zero whenever two pairs share a map, since those estimates are then built from overlapping spectra. Retaining these shared-map terms is what makes the network solve a genuine generalized least-squares problem rather than an inverse-variance-weighted average of independent edges. Any other reference gives the same angle differences.

We compare the network solution with the MK map angles after projecting both onto the same seven-dimensional differential subspace. Write $\hat{\bm q}_{\RelCal}$ and $\hat{\bm q}_{\MK}$ for the two projected angle vectors and $\Delta\bm q\equiv\hat{\bm q}_{\RelCal}-\hat{\bm q}_{\MK}$ for their difference, with $\bm C_{\RelCal}$ and $\bm C_{\MK}$ the covariances of each and $\bm C_{\RelCal,\MK}$ their cross-covariance. The two estimates use the same sky maps and are therefore correlated. Linearizing both estimators about the measured spectra (Appendix~\ref{app:covariance}) gives
\begin{equation}\label{eq:global_chi2}
\begin{aligned}
\chi^2_{\rm diff}&=\Delta\bm q^{\trans}\,\bm C_{\Delta q}^{-1}\,\Delta\bm q,\\
\bm C_{\Delta q}&=\bm C_{\RelCal}+\bm C_{\MK}-\bm C_{\RelCal,\MK}-\bm C_{\MK,\RelCal},
\end{aligned}
\end{equation}
with shared-data pulls $z_i=\Delta q_i/\sqrt{[\bm C_{\Delta q}]_{ii}}$. Retaining the cross terms is essential: treating the two reconstructions as independent would misstate the significance of their difference.

\section{Data and Implementation}\label{sec:data}

We use the eight \textit{Planck} \npipe{} (PR4) detector-set polarization maps at 100, 143, 217, and 353\,GHz, split A/B \citep{PlanckNPIPE}, giving 28 unordered pairs. We form pseudo-$C_\ell$ bandpowers with \NaMaster{} \citep{Alonso2019} from \textsc{HEALPix} maps \citep{Gorski2005}, over $51\le\ell\le1490$ with $\Delta\ell=20$. The primary mask retains $f_{\rm sky}=0.92$; a conservative $f_{\rm sky}=0.62$ mask is used only for the conditional common-mode check in Section~\ref{sec:anchor}. Cross-spectra between distinct detector sets avoid the leading noise auto-bias, though noise still enters the variances. At 217 and 353\,GHz the large-scale signal is foreground dominated, which gives high signal-to-noise differential calibration but also motivates the foreground check below.

\emph{Reference likelihood.} We use an independent implementation of the MK likelihood \citep{MinamiKomatsu2020b,DiegoPalazuelos2022,EskiltKomatsu2022} as the absolute reference, jointly fitting $\{\alpha_i\}$ and $\beta$. Intrinsic dust $EB$ is modeled with amplitudes $A_k$ in four multipole ranges, multiplying $\sin(4\psi_\ell)$ times the observed $EE$ spectrum; $\psi_\ell=\tfrac12\arctan(\Cl{TB}/\Cl{TE})$ is measured from 353\,GHz A$\times$B spectra and smoothed in $\ell$. The covariance is rotated by the trial angles, and the corresponding log-determinant term, $\ln\det C$, in the Gaussian likelihood normalization is retained. For the nearly full-sky mask,
\begin{equation}\label{eq:mk_beta_value}
\beta_{\MK}=0.37\pm0.11^\circ,
\end{equation}
consistent with the $\beta=0.342^{+0.094^{\circ}}_{-0.091^{\circ}}$
of \citet{EskiltKomatsu2022}, whose implementation we follow; the larger uncertainty here reflects our use of the four \textit{Planck} HFI frequencies alone, without the WMAP and LFI channels of that joint analysis. The differential comparison uses only $\dalpha_{ij}^{\MK}=\alpha_i^{\MK}-\alpha_j^{\MK}$, so its result is insensitive to the MK common mode.

\emph{Foreground $E/B$ SED check.} The cross-frequency condition from Section~\ref{sec:relative} is that dust $E$ and $B$ modes have nearly common frequency scaling. At $\ell<300$, the measured dust-dominated $C_\ell^{EE}/C_\ell^{BB}$ ratios are 1.41, 1.37, and 1.42 at 143, 217, and 353 GHz, a 3.6\% spread in power ratio. This constrains the amplitude-level mismatch between the dust $E$- and $B$-mode frequency scalings to $\eta\simeq0.018$ (defined in Appendix~\ref{app:relative}, where the resulting bias is derived), implying a cross-frequency bias below $0.01^\circ$ for these data. This validates the separable foreground approximation on the fiducial selection. It is not a substitute for end-to-end tests with spatially varying SEDs or frequency decorrelation, which should be revisited at higher multipoles and for more aggressive masks.

\section{Differential Consistency Results}\label{sec:results}

\begin{figure}[t]
\centering
\includegraphics[width=\columnwidth]{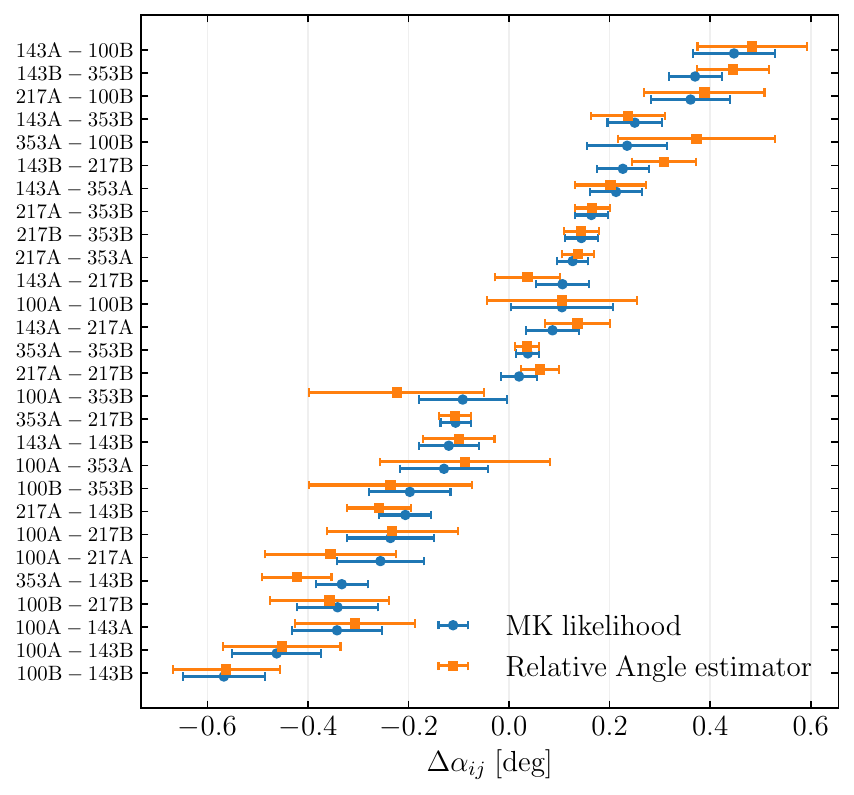}
\caption{Pairwise differential polarization angles for the 28 \textit{Planck} detector-set pair. Blue circles show the foreground-assisted MK likelihood, while orange squares show the birefringence-blind relative estimator. The orange error bars represent uncertainties on the directly estimated pairwise differences $\Delta\alpha_{ij}=\alpha_i-\alpha_j$, whereas the blue error bars represent the $1\sigma$ likelihood uncertainties of the individual map angles $\alpha_i$ from the MK analysis. The common rotation $\beta$ is absent from both reconstructions.}
\label{fig:pairwise}
\end{figure}

Figure~\ref{fig:pairwise} shows the central data comparison. The 28 pair angles track one another over the full $\pm0.5\degree$ range, with Pearson correlation $r=0.98$. The edges share maps, sky, and spectra and span only seven independent map-angle modes, so this coefficient is descriptive rather than a significance test, and we compare the reconstructions in their common network space instead.

\begin{figure}[t]
\centering
\includegraphics[width=\columnwidth]{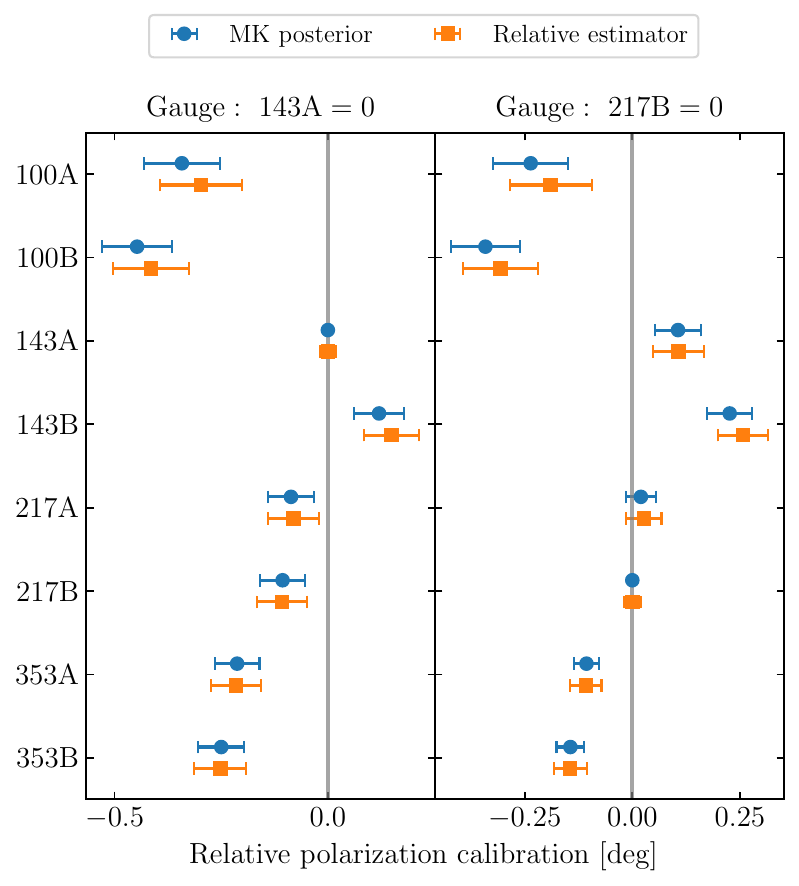}
\caption{Gauge-fixed differential calibration in two independent gauges: $\alpha_{143A}=0$ (left) and $\alpha_{217B}=0$ (right). Squares show the relative-estimator network solution and circles the MK angles after removal of their common mode. The displayed zero point changes with gauge, whereas angle differences and the network consistency statistic do not.The error bars here are uncertainties on the reconstructed map angles after the correlated pairwise measurements have been combined, whereas Figure~\ref{fig:pairwise} shows uncertainties for individual map-pair estimates.}
\label{fig:network}
\end{figure}

Figure~\ref{fig:network} shows the corresponding map-level comparison for 143A and 217B as reference maps. Both reconstructions recover the same ordering and amplitude of detector-dependent offsets. Projected onto the seven observable differential modes, their difference is
\begin{equation}\label{eq}
\chi^2_{\rm diff}=8.18\quad(7~{\rm dof}),\qquad \PTE=0.32,
\end{equation}
using the shared-data covariance of Equation~(\ref{eq:global_chi2}). The statistic is invariant under the choice of reference map, as verified using each of the eight maps in turn. All shared-data pulls are below $1.8\sigma$; the largest is $z=+1.8$ for 143B, both 217\,GHz maps have $|z|<0.4$, and the same-frequency 217A--217B pair has $z=-0.1$. No statistically significant differential inconsistency is present.

This is an important consistency result, though not an independent-data confirmation, since both reconstructions use the same sky realization. The detector-dependent pattern favored by the foreground-assisted likelihood is also recovered by a birefringence-blind estimator that never invokes that likelihood's foreground-$EB$ model, which strengthens confidence in the calibration foundation of the MK-inferred $\beta$ without testing its absolute zero point.

The comparison is sensitive enough to be useful for future analyses. A miscalibration of a single map would be detected at 95\% confidence at $0.02$--$0.03\degree$ for the high-signal 217 and 353\,GHz maps, rising to $0.1$--$0.2\degree$ at 100\,GHz (Appendix~\ref{app:power}). These are thresholds on a discrepancy between the two estimators, not on a physical angle shift common to both. Fitting one offset at a time to the observed residuals gives a largest reduction of only $\Delta\chi^2=2.2$, for 143B, insignificant after the eight-map look-elsewhere factor.

\section{Conditional Common-Mode Reconstruction}\label{sec:anchor}

The differential test cannot determine the network zero point. Given total rotations $\bm\theta$ and an absolute calibration vector $\bm\alpha^{\rm anc}$, the residual $\bm y=\bm\theta-\bm\alpha^{\rm anc}$ should equal $\beta\bm1$. The generalized-least-squares common mode is
\begin{equation}\label{eq:anchored}
\hat\beta=\frac{\bm1^{\trans}\bm C_y^{-1}\bm y}{\bm1^{\trans}\bm C_y^{-1}\bm1},
\qquad
\sigma_\beta^2=\big(\bm1^{\trans}\bm C_y^{-1}\bm1\big)^{-1},
\end{equation}
where $\bm C_y$ is the covariance of the residual vector $\bm y$. Since $\bm\theta$ and $\bm\alpha^{\rm anc}$ are both derived from the same maps, $\bm C_y=\bm C_\theta+\bm C_{\rm anc}-\bm C_{\theta,\rm anc}-\bm C_{\rm anc,\theta}$, with the total-rotation and anchor covariances obtained from their respective fits and the cross terms evaluated by the linearization of Appendix~\ref{app:covariance}. Thus relative spectra determine the differential pattern, while an external anchor is required for an absolute $\beta$.

No external sub-$0.1\degree$ anchor is available for \textit{Planck}, so we adopt the common mode of the MK likelihood itself. The result is therefore a conditional closure quantity rather than an independent determination of $\beta$, but it remains a useful subset check: the anchor comes from the full-channel MK calibration, while the total rotations are refit from the CMB-dominated 100 and 143\,GHz maps alone.

\begin{deluxetable}{lcc}[t]
\tablecaption{Conditional anchored common-rotation reconstruction.\label{tab:anchor}}
\tablehead{\colhead{Analysis} & \colhead{$\beta$ [deg]} & \colhead{$\sigma_\beta$ [deg]}}
\startdata
MK, nearly full sky & 0.37 & 0.11\\
Anchored, nearly full sky & 0.37 & 0.12\\
Anchored, $f_{\rm sky}=0.62$ Galactic mask & 0.40 & 0.13\\
\enddata
\tablecomments{Both anchored rows use the calibration anchor inferred on the nearly full-sky selection; only the total-rotation fit changes.}
\end{deluxetable}

Table~\ref{tab:anchor} summarizes the results. The anchored values agree with the MK result, and changing the mask shifts the central value by only $0.04\degree$; we assign no formal tension, since the two selections share both the anchor and most of the sky. Varying the minimum multipole shows no significant degradation up to $\ell_{\min}\simeq200$; the uncertainty increases by $5\%$ at $\ell_{\min}\simeq500$ and $20\%$ at $\ell_{\min}\simeq700$, with further degradation at higher $\ell_{\min}$. These checks are secondary to the differential result, and mainly illustrate how the network framework would accept an external anchor in a future experiment.

\section{Discussion and Recommendations}\label{sec:conclusions}

We developed a birefringence-blind differential calibration estimator and applied it for the first time as an explicit consistency test of a foreground-assisted cosmic-birefringence analysis. The common rotation cancels identically, leaving the detector-dependent pattern $\alpha_i-\alpha_j$. For the CMB and same-frequency pairs the cancellation is exact; for the cross-frequency \textit{Planck} edges, the measured dust $E/B$ stability bounds the separable-foreground bias below $0.01\degree$ on the fiducial selection. The network comparison with the MK solution gives $\chi^2=8.18/7$ ($\PTE=0.32$), with every shared-data pull below $1.8\sigma$.

Because the two reconstructions use the same maps, this is not an independent-data confirmation and it does not establish the absolute value of $\beta$. It is instead an independent estimator, with a different foreground dependence, for the observable part of the calibration. Its agreement with the MK angles shows that the detector-dependent structure is supported by the polarization data itself and is not an artifact of the absolute foreground-$EB$ model, which is the immediate scientific use of the method.

We recommend reporting this differential validation alongside future $EB$-based birefringence measurements: the network $\chi^2$ and PTE, the shared-data pulls, and the single-map discrepancy thresholds. A complete validation program should also test the analytic covariance and the cross-frequency approximation against end-to-end simulations with spatially varying dust SEDs, frequency decorrelation, bandpass differences, and realistic noise and systematics. Such tests lie beyond this Letter and become more important at higher multipoles and for more aggressive masks.

The differential network does not fix the absolute zero point. For \textit{Planck}, external polarized calibrators remain limited at the precision relevant for $\beta$ \citep{Ritacco2024}. In a future experiment, a sub-$0.1\degree$ anchor from a dedicated calibrator or a separately calibrated overlapping telescope can be combined with Equation~(\ref{eq:anchored}) to obtain an absolute birefringence measurement. The same validation layer is therefore well suited to forthcoming BICEP/Keck, SPT-3G, Simons Observatory, and LiteBIRD analyses \citep{BICEPKeck:2026inv, SPT-3G:2014dbx,SimonsObservatory2019,LiteBIRD2023}; the Simons Observatory SAT$\times$LAT configuration provides a natural example of the required relative-calibration geometry \citep{LonappanSATxLAT2026}.

\begin{acknowledgments}
We thank Eiichiro Komatsu, Patricia Diego-Palazuelos, Johannes R. Eskilt, and the referee for their detailed and constructive comments, which substantially improved this Letter. We thank the \textit{Planck} Collaboration for making the \npipe{} polarization maps and associated products publicly available. This work made use of \texttt{NumPy} \citep{Harris2020numpy}, \texttt{SciPy} \citep{Virtanen2020scipy}, \texttt{Matplotlib} \citep{Hunter2007matplotlib}, \texttt{healpy}/\textsc{HEALPix} \citep{Gorski2005}, \NaMaster{} \citep{Alonso2019}, and \texttt{emcee} \citep{ForemanMackey2013emcee}. The analysis was developed using computational resources at NERSC.
\end{acknowledgments}

\section*{Data Availability}
The full analysis pipeline, including the relative-calibration estimator, the network reconstruction, and the analytic shared-data covariance, is publicly available at \url{https://github.com/antolonappan/cosmic_birefringence}. The version used in this work is archived at \citet{cosmic_birefringence_zenodo}.

\appendix
\section{The Relative-Angle Identity and the Cancellation of $\bar D^{EB}_{ij}$}\label{app:relative}

Throughout this appendix, barred fields and spectra are pre-rotation quantities, while unbarred fields and spectra are the observed quantities after rotation. Substitution of the rotation in Section~\ref{sec:relative} therefore gives
\begin{align}
S_{ij,\ell}&=\bar S_{ij,\ell}\cos[2(\theta_i-\theta_j)]-\bar D^{EB}_{ij,\ell}\sin[2(\theta_i-\theta_j)],\\
D^{EB}_{ij,\ell}&=\bar S_{ij,\ell}\sin[2(\theta_i-\theta_j)]+\bar D^{EB}_{ij,\ell}\cos[2(\theta_i-\theta_j)],
\end{align}
where $\bar S_{ij,\ell}=\Cl{\bar E_i\bar E_j}+\Cl{\bar B_i\bar B_j}$ and $\bar D^{EB}_{ij,\ell}=\Cl{\bar B_i\bar E_j}-\Cl{\bar E_i\bar B_j}$. The common $\beta$ cancels for any angle, and $\bar D^{EB}_{ij,\ell}=0$ gives Equation~(\ref{eq:tan_relation}).

For the CMB, $\bar X_{i,\ell m}=b_i(\ell)X_{\ell m}$, so $\bar D^{EB,\rm CMB}_{ij,\ell}=b_i(\ell)b_j(\ell)(\Cl{BE}-\Cl{EB})=0$ even for nonzero cosmological $\Cl{EB}$. For a separable polarized foreground,
\begin{equation}\label{eq:Dij_fg}
\bar D^{EB,\rm fg}_{ij,\ell}=b_i(\ell)b_j(\ell)\,\Cl{EB,\rm fg}\big[a_B(\nu_i)a_E(\nu_j)-a_E(\nu_i)a_B(\nu_j)\big].
\end{equation}
The bracket vanishes for same-frequency pairs and for a common $E/B$ SED. A residual $\bar D^{EB}_{ij,\ell}\neq0$ produces, to first order, a bias whose magnitude is
\begin{equation}
\left|\delta(\dalpha_{ij})\right|\simeq\tfrac12\left|\frac{\bar D^{EB}_{ij}}{\bar S_{ij}}\right|
\lesssim\tfrac12 f_{\rm dust}\eta\epsilon_{EB}.
\end{equation}
Using $f_{\rm dust}\lesssim0.6$, the amplitude-level mismatch $\eta\simeq0.018$ inferred from the 3.6\% power-ratio spread, and $\epsilon_{EB}\sim0.03$ \citep{Clark2021,Vacher2023}, gives $|\delta(\dalpha)|\lesssim0.01\degree$ for the most exposed 353\,GHz edges. End-to-end dust simulations with spatially varying $E/B$ SEDs and frequency decorrelation remain an important future validation \citep{Clark2021,HerviasCaimapo2024}.

\section{Shared-Data Covariance}\label{app:covariance}

The relative angles are linearized functions of the bandpowers through Equation~(\ref{eq:pair_chi2}), while the MK angles are linearized through the response matrix of the joint likelihood. Propagating the analytic Gaussian bandpower covariance gives $\bm C_{\RelCal}$, $\bm C_{\MK}$, and the cross blocks in Equation~(\ref{eq:global_chi2}). The cross terms are positive because both reconstructions use the same sky realization; they are retained in all quoted pulls and PTEs. The anchored result includes the corresponding anchor--rotation covariance. End-to-end Monte Carlo validation of this approximation is left to future work.

\section{Detection Thresholds and Single-Map Decomposition}\label{app:power}

Injecting an offset into one map and propagating it through Equation~(\ref{eq:global_chi2}) gives 95\% single-map discrepancy thresholds of $0.02$--$0.03\degree$ for the dust-dominated 353 and 217\,GHz maps and $0.1$--$0.2\degree$ for 100\,GHz. The latter are broader because of lower polarized signal and cross-polarization response \citep{DiegoPalazuelos2023}. Fitting one offset at a time to the observed residuals gives a largest reduction of $\Delta\chi^2=2.2$ (143B), followed by 2.0 (217A) and 1.8 (217B), all insignificant after the eight-map look-elsewhere factor.

\bibliography{misc/ref}
\bibliographystyle{aasjournalv7}

\end{document}